\documentclass{iau}
\usepackage{graphicx}

\def\mso{{\rm M}_\odot}

\def\simgr{\,\hbox{\hbox{$ > $}\kern -0.8em \lower 1.0ex\hbox{$\sim$}}\,}
\def\simle{\,\hbox{\hbox{$ < $}\kern -0.8em \lower 1.0ex\hbox{$\sim$}}\,}

\title[Magnetic Fields in Stars] 
{Magnetic Fields in Stars: Origin and Impact}

\author[N. Langer]   
{N. Langer}

\affiliation{Argelander-Institut f\"ur Astronomie, Universit\"at Bonn}

\pubyear{2014}
\volume{302}  
\pagerange{119--126}
\jname{Magnetic Fields Throughout Stellar Evolution}
\editors{P. Petit, ed.}
\begin{document}

\maketitle

\begin{abstract}
Various types of magnetic fields occur in stars: small scale fields, large scale fields,
and internal toroidal fields. While the latter may be ubiquitous in stars due to differential
rotation, small scale fields (spots) may be associated with envelop convection in all low and
high mass stars. The stable large scale fields found in only about 10\% of intermediate mass and 
massive stars may be understood as a consequence of dynamical binary interaction,
e.g., the merging of two stars in a binary. We relate these ideas to magnetic fields in white dwarfs
and neutron stars, and to their role in core-collapse and thermonuclear supernova explosions.
\keywords{Stars, magnetic fields, stellar evolution, supernovae}
\end{abstract}

\firstsection 
\section{Introduction}
Magnetic fields play a vital role in all stages of stellar evolution.   
This is already true during star formation. The magnetic
support in collapsing molecular cloud cores is fundamentally affecting the 
fragmentation process (Price \& Bate 2007). Later-on, during the accretion process,
magnetic fields provide the required viscosity to bring in mass and 
to remove surplus angular momentum (Donati et al. 2007).

In this paper, we investigate the role of magnetic fields in stars once they are born.
In order to do so, we distinguish various types of magnetic fields.  
First, it is useful to distinguish stable from dynamo fields. 
As stable fields we consider those which have a decay time of the order of 
the stellar life time or more, and which therefore do not need a dynamo
action to continuously replenish them. Braithwaite \& Spruit (2004) 
showed that combined toroidal-poloidal magnetic fields can survive in the radiative envelopes
of stars for a long time, which they suggested to exist in magnetic A\,stars and in magnetic 
white dwarfs.

Other magnetic field geometries have so far been found unstable, e.g., such fields are
expected to decay on their Alfv\'en time scale (e.g., Tayler 1973). However,
inherently unstable field configurations may be present in stars over long time scales,
if a dynamo process is continuously regenerating the field (Brandenburg \& Subramanian).
It may be expected that this regeneration process leads to some time variability.
The prime example may be the Solar magnetic field, which is produced by a so called
$\alpha\Omega$-dynamo, where the B-field is generated by an interplay between the differential
rotation, which winds up poloidal field and generates toroidal field, and the $\alpha$-effect,
which generates a poloidal field from a toroidal one (R\"udiger et al. 2013).

While only stable and dynamo fields are long-lived and thus accessible to observations, 
there is some evidence for intermittent fields playing a role as well (Langer 2012).
In particular during dynamical stellar merger events, which are suspected to lead to stable fields
in the merger product (see below), the fields during the merger event itself are 
thought to be significantly stronger than thereafter. This strong intermittent component
may be responsible for a removal of a large fraction of the angular momentum during the merging process.

From the observational perspective, it is also useful to distinguish various types of fields.
For once, there are large scale fields, i.e. fields where the length scale over which 
local field maxima occur at the stellar surface is comparable to the size of the star itself.
A classical example is a dipole field, i.e. a field which has only two points of maximum field 
strength at the stellar surface, which are located at different sides of the star.  
Dipole fields which have their magnetic axis inclined to the axis of rotation are in fact common
amongst intermediate mass and high mass stars, although somewhat more complicated but still
large scale field geometries occur as well.

In contrast to the large scale fields are small scale fields, for which the length scale of significant
field variation is small compared to the stellar radius. An example for this are the
Solar sunspots. The Sun also shows that stellar magnetic fields can have various components,
as the small scale sunspots with field strengths of the order of 1000\,G, and the 
global Solar dipole field with a strengths of about 1\,G.  

Finally, from the observational perspective, we want to distinguish toroidal magnetic fields 
as a third type, since toroidal fields are essentially hidden from direct observations.
Still, they may strongly influence the evolution of stars, and may thus produce indirect evidence
of their existence.    

\section{Toroidal fields: ubiquitous?}

\begin{figure}[b]
\begin{center}
 \includegraphics[width=3.4in,angle=-90]{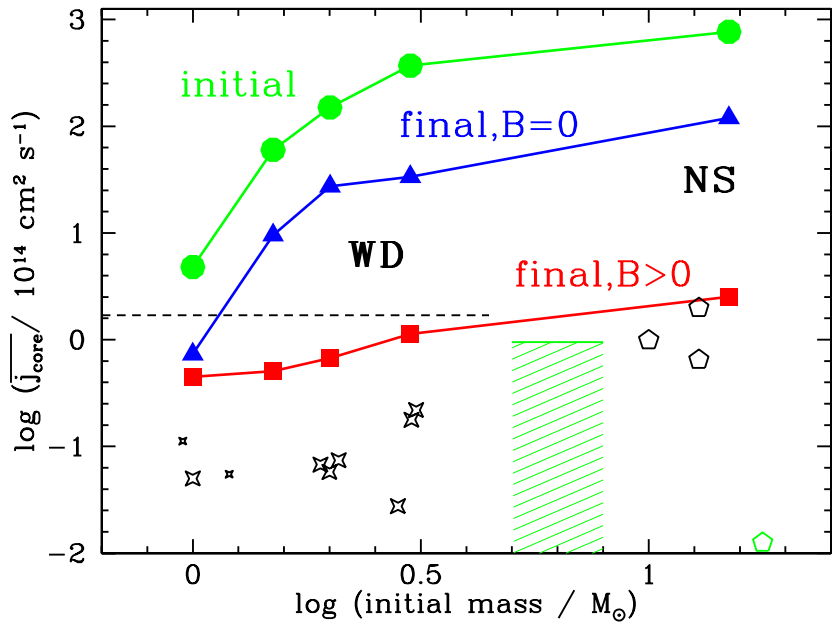} 
 \caption{
Average core specific angular momentum versus initial mass for
the low and intermediate mass models of Suijs et al. (2008)
and for the 15$\mso$ model of Heger, Woosley \& Spruit (2005)
evolving from the zero age main sequence to their end stages (full drawn lines).
The upper line corresponds
to the initial models. Filled triangles mark the final models
of the non-magnetic sequences, and filled squares  the final models
of the magnetic sequences. The dashed horizontal line indicates the
spectroscopic upper limit on the white dwarf spins obtained by Berger
et al. (2005). Star symbols represent asteroseismic measurements from
ZZ Ceti stars, where smaller symbols correspond to less certain measurements.
The green hatched area is populated by magnetic white dwarfs. The three black open
pentagons correspond to the youngest galactic neutron stars, 
while the green pentagon is thought to roughly correspond
to magnetars. See Suijs et al. (2008) for details. 
}
\end{center}
\end{figure}

Spruit (2002) has suggested that a dynamo process can operate in differentially rotating
radiative stellar envelopes. The main component of the produced magnetic field is toroidal,
which is thought to counteract the differential rotation by producing a torque which
transports angular momentum against the angular momentum gradient.
While the model of Spruit has been criticized (Zahn, Brun \& Mathis 2007),
the main effect has been confirmed in simplified MHD models (Braithwaite 2006).

While the toroidal fields are not directly observable, there are currently two lines of
observational evidence in their support. First, the nearly rigid rotation in the Sun
beneath the Solar convection zone has been reproduced by Eggenberger et al. (2005) relying on the
Spruit mechanism. Second, Mosser et al. (2012) found through oscillation measurements in
a large sample of pulsating giants that the red giant cores rotate much slower than
expected when only non-magnetic angular momentum transport processes are taken into account.
While they claim that their results agree with the slow observed spins of white dwarfs,
the evolutionary models of Suijs et al. (2007) which include angular momentum transport
through the Spruit mechanism predict indeed white dwarf and neutron star spin periods 
which are close to the observed values (Fig.~1). While it can not be excluded that  
non-magnetic transport processes like gravity waves (Talon \& Charbonnel 2008) could
also reproduce the observational constraints, the results quoted above may speak in favor of the
Spruit mechanism.  

The only prerequisite of the Spruit mechanism is differential rotation.
In case the Spruit mechanism works as expected, we may than conclude that toroidal
magnetic fields are present in {\em all} stars,
perhaps with the exception of stars with strong internal large scale fields, which 
perhaps rotate as rigid bodies, 
and of fully convective stars --- where the Spruit mechanism may be overpowered by the 
predominance of convection.
 
\section{Low mass stars}

We define low mass stars as such stars which have convective envelopes during core
hydrogen burning. The Sun is a low mass star. Magnetic activity in low mass stars
is investigated since may decades, and many papers in these proceedings give the
status of the current research. We therefore restrict ourselves here to address the magnetic 
fields in low mass stars just for comparison to those in more massive stars.

The present conclusion is that {\em all} low mass stars show magnetic fields. I.e., the presence 
of a convective envelope is sufficient to develop a field. While rotation is a necessary ingredient
to the $\alpha\Omega$-dynamo, and faster rotators tend to show higher magnetic activity,
even slow rotators as our Sun possess an appreciable magnetic field. It is clear that
these fields lead to an efficient angular momentum loss over the lifetime of these stars,
to the extent that the their spin rate is a function of their age.

\section{Intermediate mass stars}

Intermediate mass stars prove mostly to be non-magnetic. Only about 10\% of the
core hydrogen burning stars in this mass range
show a strong large scale field of more than a few hundred Gauss, while the remaining
90\% appear to have fields which are weaker than about one Gauss. 
While the fields in the magnetic fraction of intermediate mass
main sequence stars appear, partly, to have a rather complex morphology
(Donati et al. 2006), their structure appears to be simple compared to the fields of low mass stars
(Donati \& Landstreet 2009).

The magnetic intermediate mass stars are generally slow rotators. I.e., while the field strengths
of low mass stars are larger for faster rotators, the situation is almost the reverse for 
intermediate mass stars. While their cores are convective and could produce a magnetic
field deep down (Brun et al. 2005), it appears
unlikely that this field is transported deeply into the radiative envelope
or even to the stellar surface (Charbonneau \& MacGregor 2001, MacGregor \& Cassinelli 2003,
MacDonald \& Mullan 2004).
Since also their envelops are radiative, it appears not to be
possible to explain their magnetic fields through a dynamo process.  

As mentioned above, Braithwaite \& Spruit (2004) found stable magnetic field configurations
to be able to exist in these stars. However, while the suggested field geometries are
quite compatible to those observed, this does not allow any conclusion on the
origin of these fields. We return to this question on Sect.~6.

\section{Massive stars}

Evidence is accumulating that massive main sequence stars show both, the small scale
fields produced by convective envelop dynamos, and the large scale stable fields 
just as they are observed in the intermediate mass stars.

\subsection{Small-scale fields}

\begin{figure}[b]
\begin{center}
 \includegraphics[width=3.4in, angle=-00]{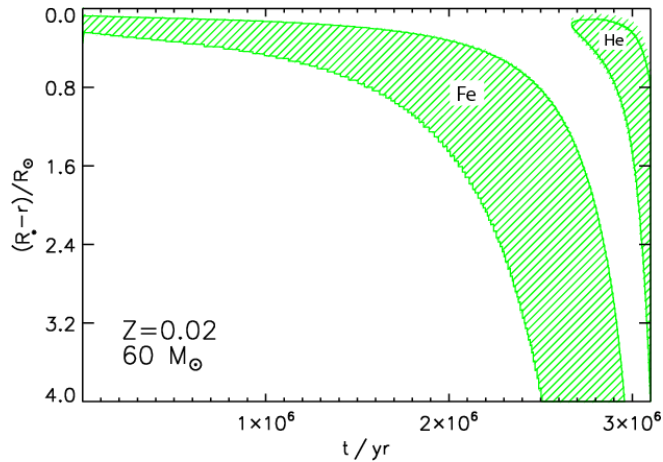}
 \caption{
Evolution of the radial extent of the subsurface helium and iron convective regions (hatched)
as function of time, from the zero age main sequence to roughly the end of core hydrogen burning,
for a 60  $~{M}_\odot$ star (Cantiello et al. 2009). The top of the plot represents the stellar surface. Only the upper
4$~{R}_\odot$ of the star are shown in the plot, while the stellar radius itself increases
during the evolution. The star has a metallicity of Z=0.02, and its effective temperature decreases
from 48 000 K to 18 000 K during the main sequence phase.
}
   \label{fig1}
\end{center}
\end{figure}

Cantiello et al. (2009) pointed out that massive main sequence stars have convective envelops
which may be capable to produce observable magnetic fields. While the convection zones occur
beneath the stellar surface due to opacity peaks produced by iron and helium recombination,
their distance to the surface is so small that magnetic flux tubes can buoyantly float to the surface 
in a short time. Their spatial extent is a significant fraction of the stellar radius
(Fig.~2).   

There is multiple observational evidence for the existence of these sub-surface convection zones.
First, Cantiello et al. (2009) showed that the predicted dependence of the kinematic signature
of these zones at the stellar surface on stellar mass, surface temperature and metallicity
agrees with the observations of micro-turbulence as determined from spectroscopic measurements of
a large number of O~and early B~stars in the Galaxy and the Magellanic Clouds. Secondly,
stochastically excited pulsations have been measured in several massive main sequence stars
(cf., Belkacem et al. 2009). And thirdly, the velocity field induced by the sub-surface convection
may lead to a clumping of the hot star winds very near to their surface, in agreement with
observations (Cantiello et al. 2009,  Sundqvist \& Owocki 2013).

Finally, and most relevant in the present context, magnetic spots at the surface
of massive main sequence stars as predicted by Cantiello et al. (2009) may provide an
explanation of the discrete absorption components (DACs) in their UV spectra lines.
The DACs phenomenon appears to be best explained by assuming a disturbance of the
radiation driven wind of the star due to hot spots at its surface (Cranmer \& Owocki
1996). As DACs are ubiquitous in O stars (Howarth \& Prinja 1989), the tentative implication is
that so are their small scale magnetic fields.

\subsection{Large-scale fields}

\begin{figure}[b]
\begin{center}
 \includegraphics[width=3.4in,angle=-90]{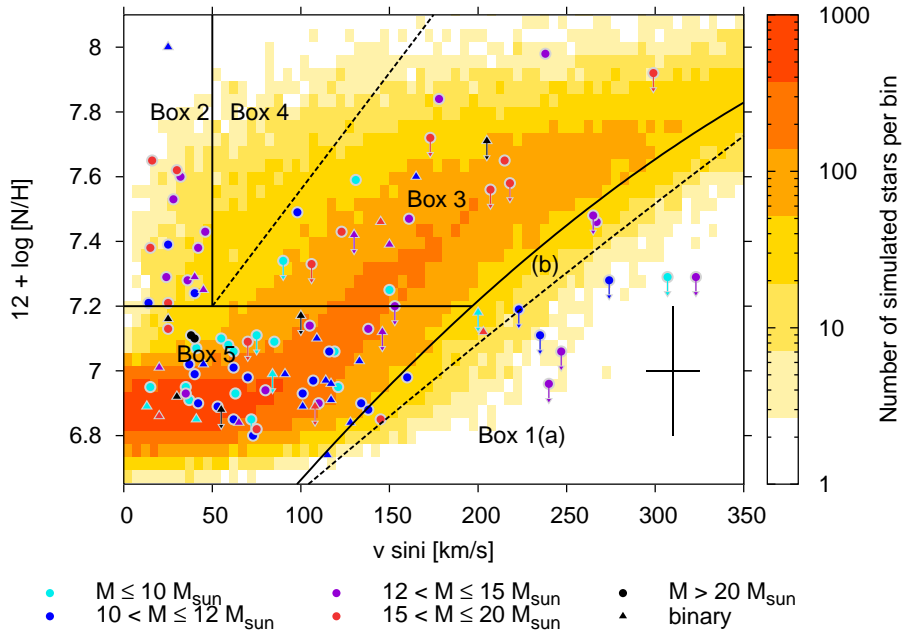}
 \caption{
Hunter diagram for LMC early B~type stars from the VLT-FLAMES Survey of Massive Stars
(symbols), showing projected rotational velocity against their nitrogen surface
abundance.
Single stars are plotted as circles, radial velocity variables as triangles.
A population synthesis simulation based on single star evolution models with rotational
mixing (Brott et al. 2011a) is shown as a density plot in the background (Brott et al. 2011b).
The color coding corresponds to the number of predicted stars per pixel.
The cross in the lower right corner shows the typical error on the observations.
}
\end{center}
\end{figure}


In recent years, the evidence has been growing that concerning large scale stable fields,
the massive stars behave essentially as their intermediate mass counterparts. 
The magnetic fraction of OB stars has been determined to be of the order of 10\%
(Grunhut \& Wade 2012),
and the magnetic topologies are similar to those found in intermediate mass stars,
with a predominance of highly inclined magnetic dipoles or low-order multipoles. 

Indirect evidence for a magnetic fraction of massive stars comes from the Hunter diagram
of LMC early B-stars (Fig. 3), which shows that 15\% of them are nitrogen-rich slow rotators
(Hunter et al. 2008, Brott et al. 2011b). While this group of stars is not reproduced from 
models of rotating single stars (Brott et al. 2011a), a magnetic field would allow to 
explain their slow rotation. Morel et al. (2008) identified a similar population in
our Galaxy, and showed that a large fraction of these objects does indeed show a magnetic field. 
Further evidence is provided by Dufton et al. (2013), who showed that the velocity distribution
of LMC early B-type stars is bimodal, with $\sim$20\% of them rotating with values below
100\,km/s, very reminiscent of the situation in the A-type stars (Zorec \& Royer 2012).

\section{The origin of stable large-scale fields}


Two ideas are pursued to understand the origin of the stable, large-scale magnetic fields 
in intermediate and massive stars. One is that the field inside the main sequence stars
is a relic of the interstellar field present in the molecular cloud at the time when
it formed the stars (e.g., Mestel 2001, Moss 2001). 
While it may appear difficult to understand how the memory of the pristine B-field is preserved
when matter is funneled through the MRI-driven accretion disc, this {\em fossil field}
hypothesis has the basic problem to explain why it is $\sim$10\% of the stars that obtain
a large scale field in this way. 

Alternatively, it has been postulated that the magnetic field as been acquired by the
star in an earlier phase of its evolution, i.e., not during its formation. 
Confusingly, fields according to this hypothesis are also sometimes called {\em fossil}.
Since the event with the largest appeal for field generation is strong close binary 
interaction, we speak here of {\em binary-induced fields}. 
The general idea that magnetic fields are generated through stellar mergers is
supported by the dearth of close companions
to magnetic main sequence stars at intermediate and high mass (Carrier et al. 2002).

Several people have suggested that strong binary interaction, in particular stellar merger, 
can result in the generation of strong, stable magnetic fields. Ferrario et al. (2009)
and Tutukov \& Fedorova (2010)
suggested to explain the magnetic intermediate mass and massive stars through
pre-main sequence mergers, while Tout et al. (2008) argued that the fields in
magnetic white dwarfs may originate from white dwarf mergers.

However, there is evidence that a large fraction of the observed magnetic intermediate 
mass and massive stars are remnants of mergers between two main sequence stars.
The latest determination of O~star main sequence binary parameter distributions by
Sana et al. (2012) implies that $8^{+9}_{-4}$\% of all Galactic O~stars are
indeed the product of a merger between two main sequence stars (de Mink et al. 2014).
Furthermore, Glebbeek et al. (2013) predict a nitrogen enrichment in main sequence 
merger products which is well compatible with the values found in the magnetic
early B-stars analyzed by Morel et al. (2008; cf. Sect.~5.2).

While indeed also a significant fraction of the intermediate mass Herbig stars
is found to be magnetic (Hubrig et al. 2004, Wade et al. 2007), 
those might also be merger products on the pre-main sequence,
possibly induced by circumstellar tides as proposed by Krontreff et al. (2012).
The ratio of nitrogen-rich to nitrogen-normal magnetic massive main sequence stars
may thus give an indication of the ratio of pre-main sequence to main sequence mergers.

A stellar merging process is a most drastic event which induces strong differential rotation
on a timescale close to the dynamical timescale. Binary evolution may produce such a situation
also in some cases where the final merger is avoided in the end, i.e. during common envelope evolution
which leads to an {\em almost-merger}. Perhaps Plaskett's star, which is a very massive 
close binary just past its rapid mass transfer phase, and of which the mass gainer has been
found to host a strong magnetic field (Grunhut et al. 2013), and the Polars, 
Cataclysmic Variables with magnetic white dwarf companions (Tout et al. 2008), 
belong to this category.

It appears in any case unlikely that thermal timescale mass transfer can lead to
strong magnetic fields in the mass gainer, neither accretion during star formation, which occurs
on a similar time scale. The point is that the products are not observed to be magnetic.
Mass transfer in massive binaries is known to lead to Be stars, which may often
be seen as single stars because their companion exploded in a supernova explosion
(de Mink et al. 2013). While perhaps not all Be stars are binary products,
{\em none} of the many analyzed Be stars has been found to be magnetic by Grunhut \& Wade (2012).
Analogously, if accretion during star formation would induce strong stable fields, 
all stars should possess such fields, which is obviously not the case.

\section{Evolution}

While in the sections above, we were mostly assessing main sequence star, one may wonder
how the stable magnetic fields survive during the post main sequence evolution
of intermediate mass and massive stars. Theoretical ideas for this are scarce, since
after the main sequence, ever changing parts of the post main sequence stars become
convective. In particular, most of these stars may evolve into red giants and supergiants,
which have convective cores and deep convective envelopes and thus leave little room
for a stable magnetic field. On the other hand, there is no stage where the whole star
would become convective, and it may thus be possible that the stable field of the main
sequence stars is preserved throughout their post-main sequence evolution deep inside the star.

It has indeed been postulated that the magnetic white dwarfs --- again: about 10\% of all
white dwarfs --- are the remnants of magnetic intermediate mass
main sequence stars, as the magnetic fluxes of both are quite comparable. 
In the light of the previous section, it may be plausible 
to assume that the magnetic white dwarfs have perhaps two components, one evolving from
magnetic main sequence stars, and the other from white dwarf mergers. The latter channel
may lead to an on average larger mass of magnetic white dwarfs, while the former might
stand out by a slower-than-average spin. If magnetic white dwarfs are really merger 
products, it may be interesting to note that they will only play a minor role in
Type~Ia supernovae, if any at all. However, magnetic fields could form in double
degenerate mergers, which are thought to provide one channel towards Type~Ia explosions.

While the fraction of neutron stars which have extreme fields, the magnetars, is not
well established, it appears again compatible with an order of magnitude of $\sim 10\%$.
Also the flux freezing argument could apply. If massive main sequence stars
had a B-field of $\sim 10^4$\,G in their core --- which appears plausible as they have surface
fields of up to $\sim 10^3$\,G ---, the resulting B-field in the neutron stars would be of the order of
$10^{14}\,$G, which is two orders of magnitude larger than typical neutron star magnetic field strengths
(Ferrario \& Wickramasinghe 2006).
In contrast to the scenario by Duncan \& Thompson (1992), where the magnetar field forms from an
extremely rapidly rotating collapsing iron core, magnetars as successors of magnetic main sequence stars
would form slowly rotating neutron stars. 
This appears not only to be more compatible with the young supernova remnants
surrounding some magnetars (Vink 2008), but also with the dearth of progenitors which can
produce rapidly rotating iron cores in a high metallicity environment as our Galaxy (Yoon et al. 2006),
and would argue against magnetar-powered supernovae (Woosley 2010).
As also proposed by Duncan \& Thompson (1992), the proto-neutron stars with ordinary 
spin rates ( corresponding to $j\simeq 10^{14}\,$cm$^2$/s; see Fig.~1) may 
well produce the $10^{12}\,$G fields found in most neutron stars. 

\section{Outlook}

While all low mass stars appear to have small scale, dynamo-produced surface fields, 
the stronger the faster they rotate, only a fraction
of $\sim 10$\% of the intermediate mass stars possess strong B-field, which are large scale
and occur mostly in slowly rotating single stars.
Both types of fields may be combined in the massive stars, the small scale ones in all of the,
the large scale one again in a fraction of about 10\%.
In addition, all stars may contain internal toroidal magnetic fields induced by differential rotation,
which couples their core and envelope spins.

The hypothesis for the formation of the large scale fields
which is consistent with all currently known constraints is that of strong binary
interaction, preferentially via stellar mergers. In contrast to the fossil field hypothesis,
it makes several clear predictions.
Due to the large observational efforts currently underway,
we can expect that this topic will be settled within the next years. 

The question of the influence and survival of the large scale fields during the post-main
sequence evolution of intermediate mass and massive stars appears more difficult to answer.
Since red supergiants will likely not allow to assess this question observationally due to
their deep convective envelopes --- which may produce its own field through a dynamo process ---, 
it may be interesting to focus on blue supergiants and 
Wolf-Rayet stars. If descendants of magnetic main sequence stars evolve into long-lived
blue supergiants, a fair fraction of them might show surface magnetic fields, although
considerably weaker ones if the magnetic flux is conserved. Also some Wolf-Rayet stars may 
have magnetic main sequence stars as precursors, but a field detection in these objects
appears difficult due to their strong winds (de la Chevroti\`ere et al. 2013). 

Whether the large scale fields survive even until the formation of the compact remnant
remains an open question, although there may be more arguments in favor of this idea
than against it (cf. Sect. 7). However, it remains a challenge to produce solid theoretical predictions 
about the survival of the field during the post-main sequence evolution, 
as well as to identify observational strategies which would allow to settle the case.

\end{document}